\documentclass[10pt,final,journal]{IEEEtran}
\usepackage[utf8]{inputenc}
\usepackage[noadjust]{cite}
\usepackage{amsmath}
\usepackage{graphicx}
\usepackage{array}
\usepackage{multirow}
\usepackage{textcomp}
\usepackage{hhline}
\usepackage{url}
\usepackage{tabularx}
\usepackage{caption} 
\usepackage{colortbl}
\usepackage{amssymb}
\usepackage{acro} 
\usepackage{cite}
\usepackage{fancyhdr}
\usepackage[ruled,vlined]{algorithm2e}
\usepackage{authblk}
\usepackage{bm}
\usepackage[final]{changes}
\definechangesauthor[name={V.S.S Kandarpa}, color=blue]{SK}
\definechangesauthor[name={Alexandre Bousse}, color=magenta]{AB2}
\usepackage[normalem]{ulem}

\DeclareAcronym{CT}{
	short=CT,
	long=computed tomography,
}
\DeclareAcronym{FC}{
	short=FC,
	long=fully connected,
}
\DeclareAcronym{ACRIN}{
	short=ACRIN,
	long=American College of Radiology Imaging Network ,
}

\DeclareAcronym{GT}{
	short=GT,
	long=ground truth,
}
\DeclareAcronym{SD}{
	short=SD,
	long=standard deviation,
}
\DeclareAcronym{PReLU}{
	short=PReLU,
	long=parametric rectified linear unit ,
}
\DeclareAcronym{DUGAN}{
	short=DUG,
	long=double U-Net generator,
}
\DeclareAcronym{GAN}{
	short=GAN,
	long=generative adversarial network,
}
\DeclareAcronym{SSIM}{
	short=SSIM,
	long=structural similarity index,
}
\DeclareAcronym{MSE}{
	short=MSE,
	long=mean squared error,
}
\DeclareAcronym{RMSE}{
	short=RMSE,
	long=root mean squared error,
}
\DeclareAcronym{PET}{
	short=PET,
	long=positron emission tomography,
}
\DeclareAcronym{SPECT}{
	short=SPECT,
	long=single photon emission computed tomography,class = {in},
}
\DeclareAcronym{MRI}{
	short=MRI,
	long=magnetic resonance imaging,class = {in},
}
\DeclareAcronym{DnCNN}{
	short=DnCNN,
	long=denoising convolutional
	neural network ,
}

\DeclareAcronym{ROI}{
	short=ROI,
	long=region of interest,
	long-plural-form = regions of interest,
}

\DeclareAcronym{SNR}{
	short=SNR,
	long=signal-to-noise ratio,
}

\DeclareAcronym{MLEM}{
	short=MLEM,
	long=maximum likelihood expectation-maximization,
}		
\DeclareAcronym{MBIR}{
	short=MBIR,
	long=model-based iterative reconstruction,
}	

\DeclareAcronym{XCAT}{
	short=XCAT,
	long=extended cardiac-torso,
}

\DeclareAcronym{GPU}{
	short=GPU,
	long=graphical processing unit,
}

\DeclareAcronym{CNN}{
	short=CNN,
	long=convolutional neural network,	
}

\DeclareAcronym{CNR}{
	short=CNR,
	long=contrast-to-noise ratio,
}	
	
\DeclareAcronym{FDG}{
	short=$^{\text{18}}$F-FDG,
	long=$^{\text{18}}$F-fludeoxyglucose,
}
\DeclareAcronym{FLT}{
	short=FLT,
	long=fluorothymidine ,
}

\DeclareAcronym{SUV}{
	short=SUV,
	long=standardized uptake values,
}
	
\DeclareAcronym{4D}{
	short=4-D,
	long=four-dimensional,
	class = {out},
}
\DeclareAcronym{3D}{
	short=3-D,
	long=three-dimensional,
}
\DeclareAcronym{2D}{
	short=2-D,
	long=two-dimensional,
}	
\DeclareAcronym{FBP}{
	short=FBP,
	long=filtered-backprojection,
}
\DeclareAcronym{OSEM}{
	short=OSEM,
	long=ordered-subsets expectation-maximisation,
}

\DeclareAcronym{mumap}{
	short=${\mu}$-map,
	long=attenuation map,
	index=$\boldmu$-map,	
}

\DeclareAcronym{UBO}{
	short=UBO,
	long=\emph{Université de Bretagne Occidentale},
}

\DeclareAcronym{HU}{
	short=HU,
	long=Hounsfield unit,
}	

\DeclareAcronym{SR}{
	short=SR,
	long=Super Resolution,
}

\ifCLASSOPTIONcompsoc
	\usepackage[caption=false,font=normalsize,labelfont=sf,textfont=sf]{subfig}
\else
	\usepackage[caption=false,font=footnotesize]{subfig}
\fi

\newcommand{\boldmu}{\bm{\mu}}

\newcommand{\boldmuhat}{\bm{\hat{\mu}}}

\newcommand{\boldx}{\bm{x}}
\newcommand{\xhat}{\hat{x}}
\newcommand{\boldxhat}{\bm{\hat{x}}}

\newcommand{\boldlambda}{\bm{\lambda}}

\newcommand{\boldlambdahat}{\bm{\hat{\lambda}}}

\newcommand{\boldy}{\bm{y}}

\newcommand{\boldybar}{\bm{\bar{y}}}
\newcommand{\boldyhat}{\bm{\hat{y}}}

\newcommand{\boldL}{\bm{L}}
\newcommand{\boldP}{\bm{P}}

\newcommand{\bbR}{\mathbb{R}}

\begin{document}

\title{DUG-RECON: A Framework for Direct Image Reconstruction using Convolutional Generative Networks}

\author{V.S.S. Kandarpa$^\ast$, Alexandre Bousse, Didier Benoit and Dimitris Visvikis \thanks{All the authors are affiliated to LaTIM, INSERM, UMR 1101, \emph{Université de Bretagne Occidentale}. \\ $^\ast$ (email:~\texttt{venkatasaisundar.kandarpa@univ-brest.fr}) }

}

\maketitle

\begin{abstract}
This paper explores convolutional generative networks as an alternative to iterative reconstruction algorithms in medical image reconstruction. The task of medical image reconstruction involves mapping of projection domain data collected from the detector to the image domain. This mapping is done typically through iterative reconstruction algorithms which are time consuming and computationally expensive. Trained deep learning networks provide faster outputs as proven in various tasks across computer vision. In this work we propose a direct reconstruction framework exclusively with deep learning architectures. The proposed framework consists of three segments, namely denoising, reconstruction and super resolution. The denoising and the super resolution segments act as processing steps. The reconstruction segment consists of a novel \ac{DUGAN} which learns the sinogram-to-image transformation. This entire network was trained on \ac{PET} and \ac{CT} images. The reconstruction framework approximates \ac{2D} mapping from projection domain to image domain. The architecture proposed in this proof-of-concept work is a novel approach to direct image reconstruction; further improvement is required to implement it in a clinical setting.
\end{abstract}

\begin{IEEEkeywords}
Medical Image Reconstruction, Deep Learning, Generative Adversarial Networks
\end{IEEEkeywords}
\acresetall

\section{Introduction}

\IEEEPARstart{T}{he} use of deep learning in medical imaging has been on the rise over the last few years. It has widely been used in various tasks across medical imaging such as image segmentation  \cite{ronneberger2015u,guo2019deep,sinha2019multi,dolz2018hyperdense,hatt2018first}, image denoising \cite{kadimesetty2018convolutional,li2020sacnn,chen2017low,yang2018low}, image analysis \cite{litjens2017survey,amyar20193,cui2018artificial}. 
The utilization of deep learning for image reconstruction is a more challenging task. Image reconstruction using deep learning corresponds to the task of mapping raw projection data retrieved from the detector to image domain data. 
One can broadly identify three different categories of approaches for the implementation of deep learning within the framework of medical image reconstruction:
\begin{itemize}
    \item[(i)] methods that use deep learning as an image processing step that improves the quality of the raw data and/or the reconstructed image \cite{gong2018pet, maier2018deep}; 
    \item[(ii)] methods that embed deep-learning image processing techniques in the iterative reconstruction framework to accelerate convergence or to improve image quality \cite{xie2019generative,kim2018penalized,gong2019iterative};
    \item[(iii)] direct reconstruction with deep learning alone without any use of traditional reconstruction methods  \cite{whiteley2019direct,zhu2018image,haeggstroem2018deeprec}.
\end{itemize}

The use of deep learning for the development of either data corrections or post-reconstruction  image based approaches (i) has shown potential to improve the quality of reconstructed images. An example of data corrections by improving the raw data through scatter correction is proposed in \cite{maier2018deep}. In this work a modified U-Net is used to estimate scatter and correct the raw data in order to improve \ac{CT} images. Denoising the reconstructed \ac{PET} images with a deep convolutional network was done in \cite{gong2018pet}. The authors used perceptual loss along with \ac{MSE} to preserve qualitative and quantitative accuracy of the reconstructed images. The network was initially trained on simulated data and then fine-tuned on real patient data. 
Despite resulting in an improvement of the reconstructed output, the above mentioned methods do not directly intervene with the reconstruction process. This can be done using the two distinct frameworks (ii) and (iii).
\par
The first one involves the incorporation of a deep neural network into an unrolled iterative algorithm where a trained neural network accelerates the convergence by improving the intermediate estimates in the iterations \cite{gong2019iterative,xie2019generative,kim2018penalized}. The paper by Gong et al. used a modified U-Net to represent images within the iterative reconstruction framework for \ac{PET} images. The deep learning architecture was trained on low-dose reconstructed images as input and high-dose reconstructed images as the output. The work by Xie et al. further extended this work by replacing the U-Net with a \ac{GAN} for image representation within the iterative framework. Kim et al incorporated a trained \ac{DnCNN} along with a novel local linear fitting function into the iterative algorithm. The \ac{DnCNN} which is trained on data with multiple noise levels improves the image estimate at each iteration. They used simulated and real patient data in their study.
The second framework, referred to as \emph{direct reconstruction}, is based on performing the whole reconstruction process (replacing the classical framework) with a deep learning algorithm, using raw data as input and reconstructed images as output \cite{whiteley2019direct,zhu2018image,haeggstroem2018deeprec}. In contrast to (i) and (ii) 
which have been extensively investigated, direct reconstruction (iii) using deep learning has been much less explored.

\par There have been to date three particular approaches that are relevant to this strategy. The deep learning architecture proposed by Zhu et al.~\cite{zhu2018image} called AUTOMAP uses \ac{FC} layers (which encode the raw data information) followed with convolutional layers.
The first three layers in this architecture are \ac{FC} layers with dimensions $2n^2$,$n^2$ and $n^2$ respectively where $n\times{}n$ is the dimension of the input image. The AUTOMAP requires the estimation of a huge number of parameters which makes it computationally intensive. Although initially developed for \ac{MRI}, AUTOMAP has been claimed to work on other imaging modalities too. Brain images encoded into sensor-domain sampling strategies with varying levels of additive white Gaussian noise were reconstructed with AUTOMAP.  Within the same concept of using \ac{FC} layers' architectures a three stage image reconstruction pipeline called DirectPET has been proposed to reduce associated computational issues  \cite{whiteley2019direct}. The first stage down-samples the sinogram data, following which a unique Radon transform layer encodes the transformation from sinogram to image space. Finally the estimated image is improved using a super resolution block. This work was applied to full body \ac{PET} images and remains the only approach that can reconstruct multiple slices simultaneously (up to 16 images). DeepPET is another approach implemented on simulated images using encoder-decoder architecture based on the neural network proposed by the visual geometric group \cite{haeggstroem2018deeprec}. Using realistic simulated data, they demonstrated a network that could reconstruct images faster, and with an image quality (in terms of root mean squared error) comparable to that of conventional iterative reconstruction techniques.  

\par In our work we explore the use of U-Net  based deep learning architectures \cite{ronneberger2015u} to perform a direct reconstruction from the sinogram to the image domain using real patient datasets. Our aim is to reduce the number of trainable parameters along with exploring a novel strategy for direct image reconstruction using generative networks. More specifically our approach consists of a three-stage deep-learning pipeline consisting of denoising, image reconstruction and super resolution segments. Our experiments included training the deep learning pipeline on \ac{PET} and \ac{CT} sinogram-image pairs. A single pass through the trained network transforms the noisy sinograms to reconstructed images. The reconstruction of both \ac{PET} and \ac{CT} datasets was considered and presented in the following sections. At this stage the work presented is a proof of concept and needs further improvement before being applied in a clinical setting.

\begin{figure}[!htbp]
	\centering
	\includegraphics[width=0.99\linewidth]{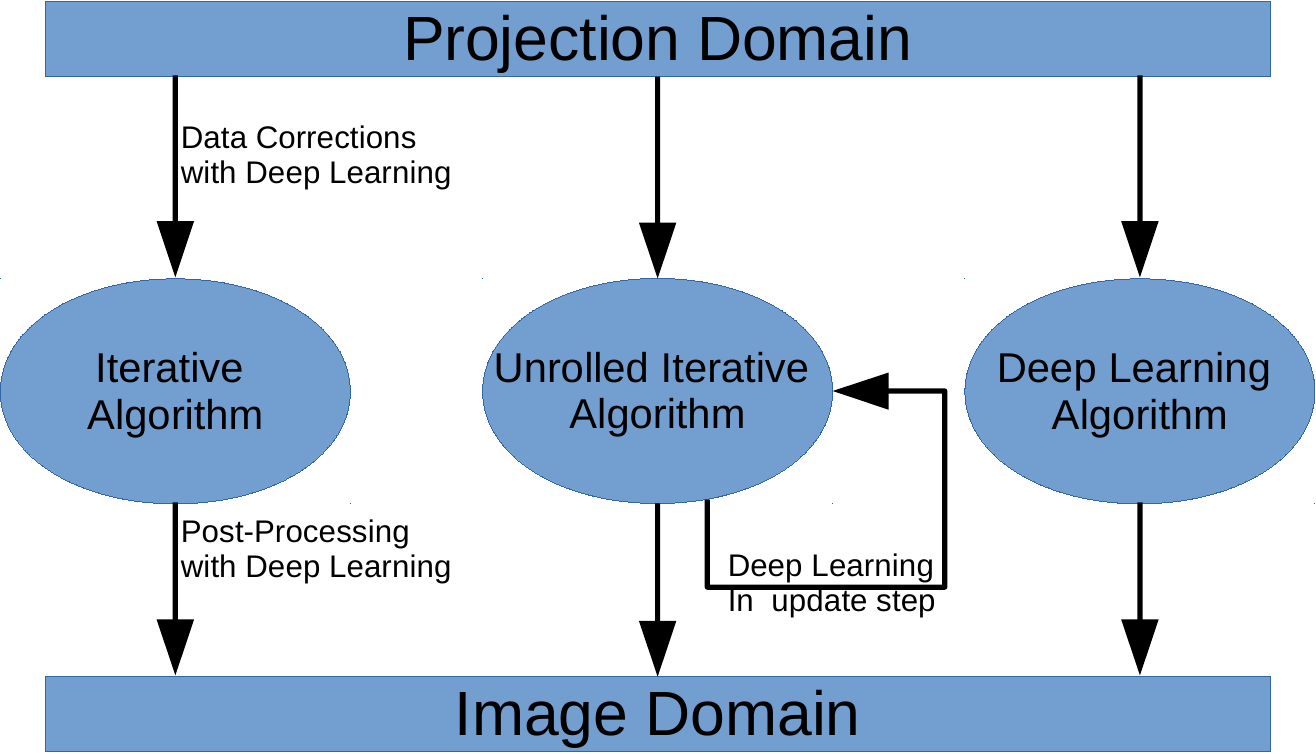}
	\caption{Deep Learning in Medical Image Reconstruction}
	\label{fig:dl}
\end{figure}

\section{Materials and Methods}

\subsection{Image Reconstruction Model}

In medical imaging, image reconstruction corresponds to the task of reconstructing an image  $\boldx \in \bbR^m$, from a scanner measurement $\boldy \in \bbR^n$, where $m$ is the number of voxels defining the image and $n$ is the number of detectors in the scanner.  In \ac{CT}, the image $\boldx = \boldmu$ corresponds to X-ray attenuation, measured by the proportion of X-rays scattered or absorbed as they pass through the object. In \ac{PET}, $\boldx = \boldlambda$ is the distribution of a radiotracer delivered to the patient by injection, and is measured through the detection of pairs of $\gamma$-rays emitted in opposite directions (indirectly from the positron-emitting radiotracer).

The measurement $\boldy$ is a random vector modeling the number of detection (photon counting) at each of the $n$ detector bins, and follows a Poisson distribution with independent entries:
\begin{equation}\label{eq:poisson}
	\boldy \sim \mathrm{Poisson}(\boldybar(\boldx))
\end{equation}    
where $\boldybar(\boldx) \in \bbR^n$ is the expected number of counts (noiseless), which is a function of the image $\boldx$. In a simplified setting, the expected number of counts in \ac{CT} is
\begin{equation}\label{eq:CT}
	\boldybar(\boldmu) = \exp (-\boldL \boldmu)
\end{equation}
where $\boldL \in \bbR^{n\times m}$ is a system matrix such that each entry $[\boldL]_{i,j}$ represents the contribution of the $j$-th image voxel to the $i$-th detector. In \ac{PET}, the expected number of counts is (also in a simplified setting) is
\begin{equation}\label{eq:PET}
	\boldybar(\boldlambda) = \boldP \boldlambda
\end{equation}
where $\boldP \in \bbR^{n\times m}$ is a system matrix such that each entry $[\boldP]_{i,j}$ represents the probability that a photon pair emitted from voxel $j$. For simplification we assume $\boldP=\boldL$ which is a reasonable assumption in a non time-of-flight setting. Image reconstruction is achieved by finding a suitable image $\boldxhat = \boldmuhat$ or $\boldlambdahat$ that approximately solves 
\begin{equation}\label{eq:pb2solve}
	\boldy = \boldybar(\boldx) \, .
\end{equation}  
\Ac{FBP} techniques (see \cite{natterer2001mathematics} for a review) can efficiently solve \eqref{eq:pb2solve} but they are vulnerable to noise as the system matrix $\boldP$ is ill-conditioned. Since the 80's, \ac{MBIR} techniques \cite{Shepp1982,fessler2000statistical} became the standard approach. They consist in iteratively approximating a solution $\boldxhat$ such that $\boldybar(\boldxhat)$ maximizes the likelihood of the measurement $\boldy$. As they model the stochasticity of the system, they are more robust to noise as compared with \ac{FBP}, and can be completed with a penalty term for additional control over the noise \cite{depierro1995}. However, they are computationally expensive as each iteration requires to compute a projection and a backprojection.  
\par
Image reconstruction with deep learning however is a data driven approach wherein there is a training and a prediction phase. Given a set of training data which is a subset of the raw data ($\boldy$) and its corresponding images ($\boldx$), a deep learning architecture learns the mapping from raw data to the image and improves this mapping through the training process. During the prediction phase a subset of raw data different from the training data serves as the input to the trained deep learning architecture. The output is a reconstructed image which is obtained on a single forward pass through the network. Hence making the reconstruction process through deep learning instantaneous as opposed to an iterative process. This makes direct reconstruction with deep learning faster and less computationally expensive than iterative algorithms.

\subsection{Deep Learning Architectures}

\begin{figure}[!htbp]
	\centering
	\includegraphics[width=0.99\linewidth]{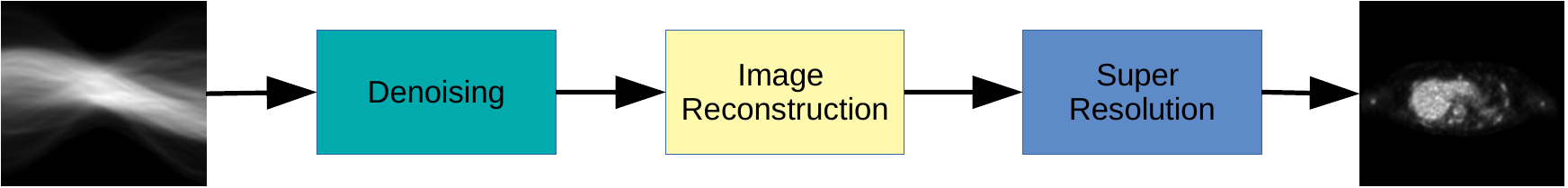}
	\caption{Proposed Deep Learning pipeline for Direct Image Reconstruction}
	\label{fig:three}
\end{figure}

As shown in Figure~\ref{fig:three}, we propose a three-stage deep learning pipeline for the task of tomographic reconstruction. In the first step the raw data (projection space) are denoised. Next the denoised sinograms are transformed to the image domain in the image reconstruction segment. The third and final segment operates in the image domain to improve the image produced after domain transformation. The following sections discuss these segments in detail.

\subsubsection{Denoising}
\begin{figure}[!htbp]
	\centering
	\includegraphics[width=0.99\linewidth]{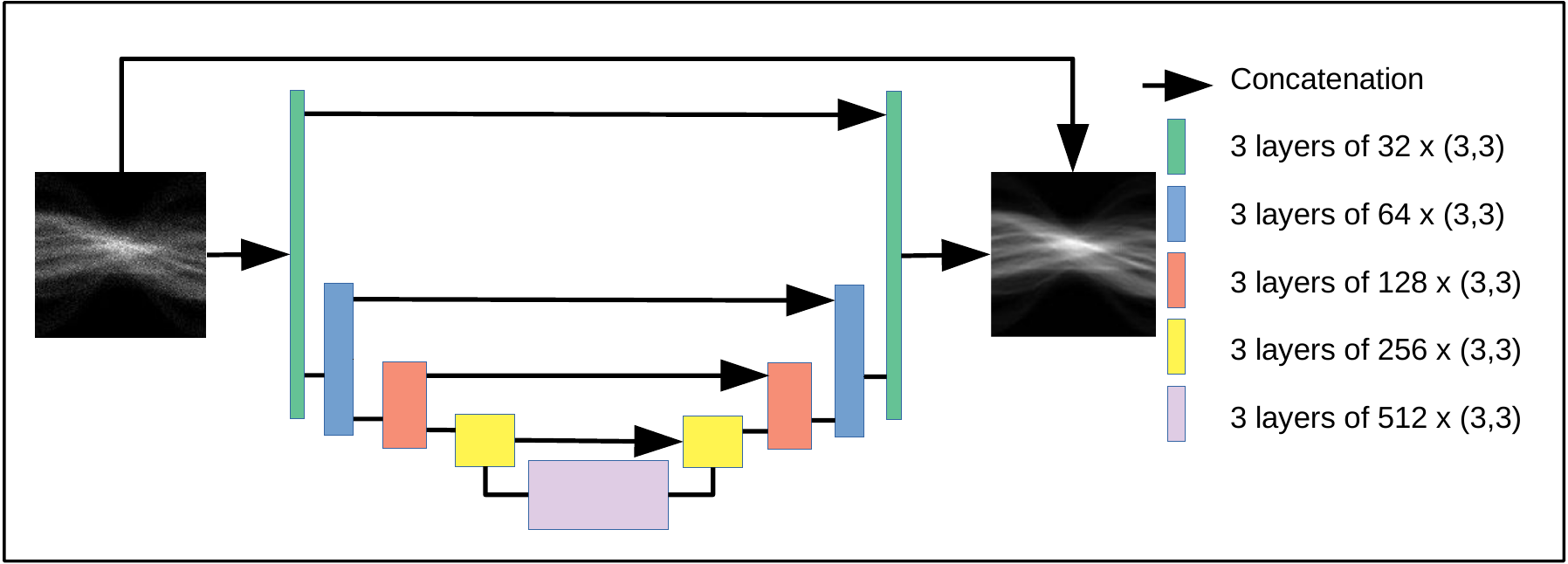}
	\caption{Representation of the denoising network. The inputs to the network were \ac{2D} grayscale slices with resolution $128\times{}128$ and the outputs were denoised sinograms.}
	\label{fig:denoise_nn}
\end{figure}

We used a modified U-Net architecture to denoise the Poisson sampled sinograms, based on the work previously carried out for ultrasound denoising \cite{perdios2018deep}. The U-Net is an encoder-decoder network which was initially implemented for segmentation but over the years its applications have broadened. As shown in Figure~\ref{fig:denoise_nn} there are increasing number of convolutions along with max pooling to arrive at an encoding of the input and then with convolutions followed by upsampling, arriving at the output with an identical dimension as the input. The important modification in the architecture mentioned in \cite{perdios2018deep} with respect to U-Net was the residual connection from the input to the final output. Perdios et al trained the denoising architecture on simulated ultrasound images so as to enhance ultrafast ultrasound imaging. This denoising architecture corresponds to the first segment in our proposed framework.
It was trained on raw data pairs, i.e., low-count and high-count sinograms, considering multiple noise levels. 
The detailed architecture is represented in Figure~\ref{fig:denoise_nn}. We defined the loss function between a true sinogram $\boldy^\star = [y_1^\star, \dots , y_n^\star]^\top \in \mathbb{R}^n$ and a prediction 
$\boldyhat = [\hat{y}_1, \dots , \hat{y}_n]^\top \in \mathbb{R}^n$ as the \ac{MSE}: 
\begin{equation}\label{eq:mse}
      \mathrm{MSE}(\boldy^\star, \boldyhat) = \frac{1}{n}   \sum_{i=1}^{n} (y^\star_i - \hat{y}_i)^2 \, ,
\end{equation}
where, $n$ is the number pixels on the sinogram, corresponding to the number of detectors in the scanner. 

\subsubsection{Image Reconstruction}
The novelty in this work is the proposed U-Net based network in contrast to previous works in direct image reconstruction using the \ac{FC} layer architectures. This design of the network draws its inspiration from conditional \ac{GAN} for image to image translation called Pix2Pix \cite{isola2017image}. The proposed network namely \ac{DUGAN} consists of two cascaded U-Nets. The first U-Net transforms the raw data to image while the second U-Net takes as input the generated image and transforms it back to the raw data. The second U-Net assesses the reconstructed image output, reiterating the relation between the sinogram and the image. This architecture differs from the Pix2Pix, which consists of a generator (U-Net like network) and a discriminator (classification convolutional network). While the generator in both architectures serves the purpose of transforming images from one domain to the other, the discriminator with regards to Pix2Pix classifies inputs as real/fake. The objective function for this architecture can be written as:
\begin{equation}
    \mathcal{L}_\mathrm{total} = \mathcal{L}_{G_{1}} + \mathcal{L}_{G_{2}} + \mathcal{L}_{G_{1}+G_{2}}
\end{equation}
where, 
\begin{equation}
    \mathcal{L}_{G_{1}} = \frac{1}{n}   \sum_{i=1}^{n} |x^\star_i - \hat{x}_i|
\end{equation}
\begin{equation}
    \mathcal{L}_{G_{2}} = \frac{1}{n}   \sum_{i=1}^{n} |y^\star_i - \hat{y}_i|
\end{equation}

$G_{1},G_{2}$ are Generator 1 and Generator 2 which predict image and sinogram respectively; $\boldx^\star$, $\boldxhat$ the true and predicted image, $\boldy^\star$, $\boldyhat$ the true and predicted sinogram respectively. $\mathcal{L}_{G_1+G_2}$ is defined similar to $\mathcal{L}_{G_2}$ with the combined architecture of $G_{1}+G_{2}$, keeping the weights of $G2$ fixed. 

\begin{figure}[!htbp]
	\centering
	\includegraphics[width=0.99\linewidth]{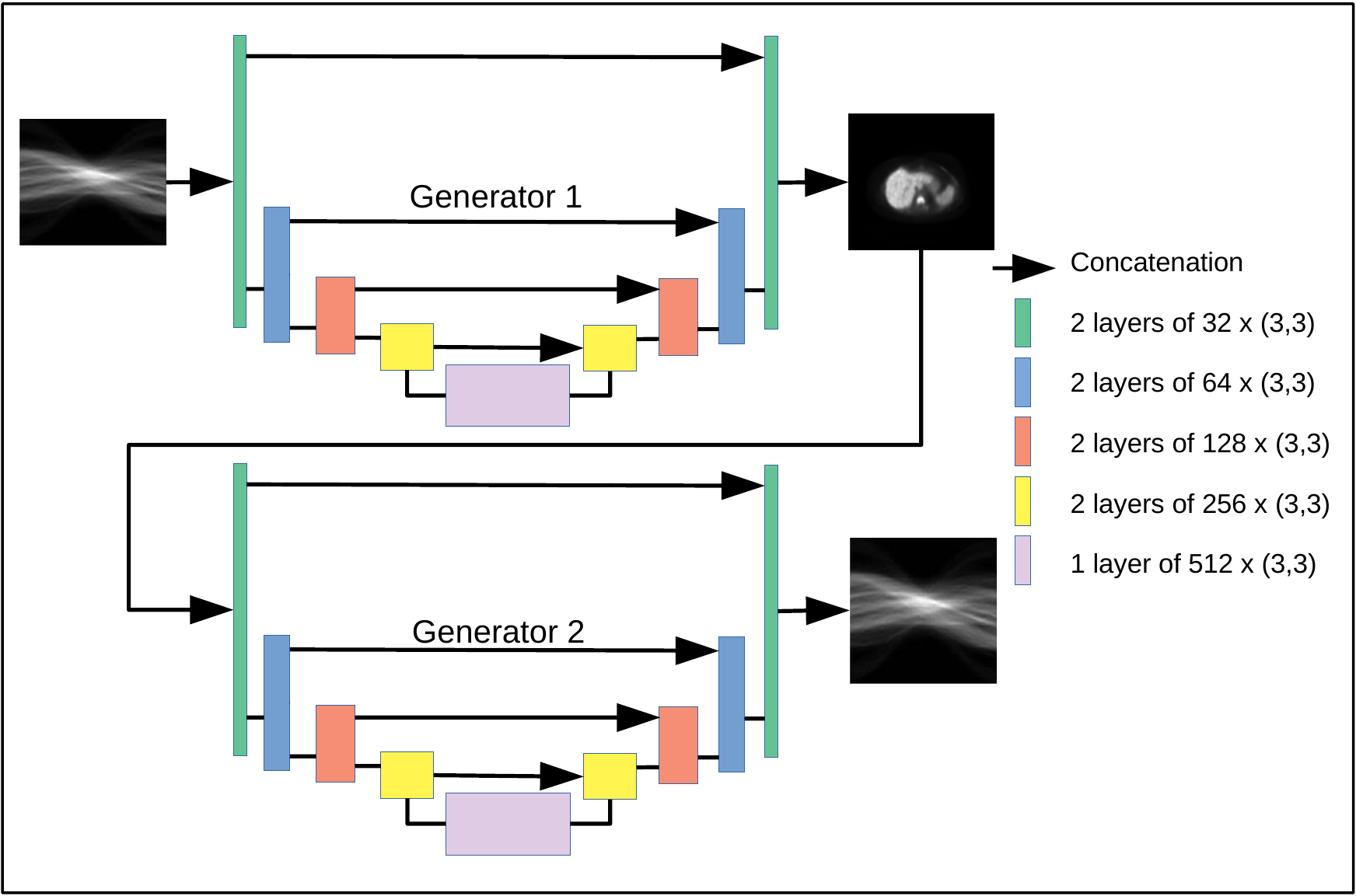}
	\caption{Representation of the \ac{DUGAN}, the image reconstruction block. This network was trained on denoised sinograms which were the outputs of the previous segment.}
	\label{fig:dugan}
\end{figure}

The architecture is represented in detail in Figure~\ref{fig:dugan}. The training for this architecture is summarized in Algorithm 1. A comparison of the trainable parameters for various segments that are used to perform the domain mapping from sinogram to image is provided in Table~\ref{table:1}, considering the AUTOMAP from \cite{zhu2018image}, the Radon inversion layer from \cite{whiteley2019direct} and the proposed architecture \ac{DUGAN} along with denoising and super resolution segments.

\begin{algorithm}
\SetAlgoLined
$M$ = number of epochs \;
$N$ = total training data (images/sinograms) \;
 \For{i = 1,2,\dots, M}{
  \For{j = 1,2,\dots,N}{
   Train $G_{1}$:  minimizing $\mathcal{L}_{G_{1}} $ \;
  }
  \For{j = 1,2,\dots,N}{
   Train $G_{2}$:  minimizing $\mathcal{L}_{G_{2}} $ \;
   }
   \For{j = 1,2,\dots,N}{
   Train combined architecture, freezing the weights of $G_{2}$: minimizing $\mathcal{L}_{G_{1}+G_{2}}$  \; }{
  }
 }
 \caption{Training the DUG}
 \label{alg:1}
\end{algorithm}

\begin{table}[h!]
	
\centering
\caption{Trainable Parameters comparison}
\label{table:1}
 \begin{tabular}{||c|c|c|c||} 
 \hline
 \textbf{Architecture} & \textbf{Input Size} & \textbf{Output Size} & \textbf{Trainable}   \\ 
                       &                     &                       & \textbf{Parameters} \\ [0.5ex] 
 \hline
 AUTOMAP & $200\times{}168$ & $200\times{}200$ & 6,545,920,000    \\ 
 \hline
 Radon Inversion Layer   & $200\times{}168$ & $200\times{}200$ & 382,259,200   \\
 ($40\times{}40$ Patch size)  &  &  & \\
 \hline
 DeepPET & $128\times{}128$ & $128\times{}128$ & 62,821,473  \\ 
 \hline
 DUG-RECON & $128\times{}128$ & $128\times{}128$ & 17,444,140  \\ 
 \hline
  \end{tabular}
\end{table}

\begin{figure}[!htbp]
	\centering
	\includegraphics[width=0.99\linewidth]{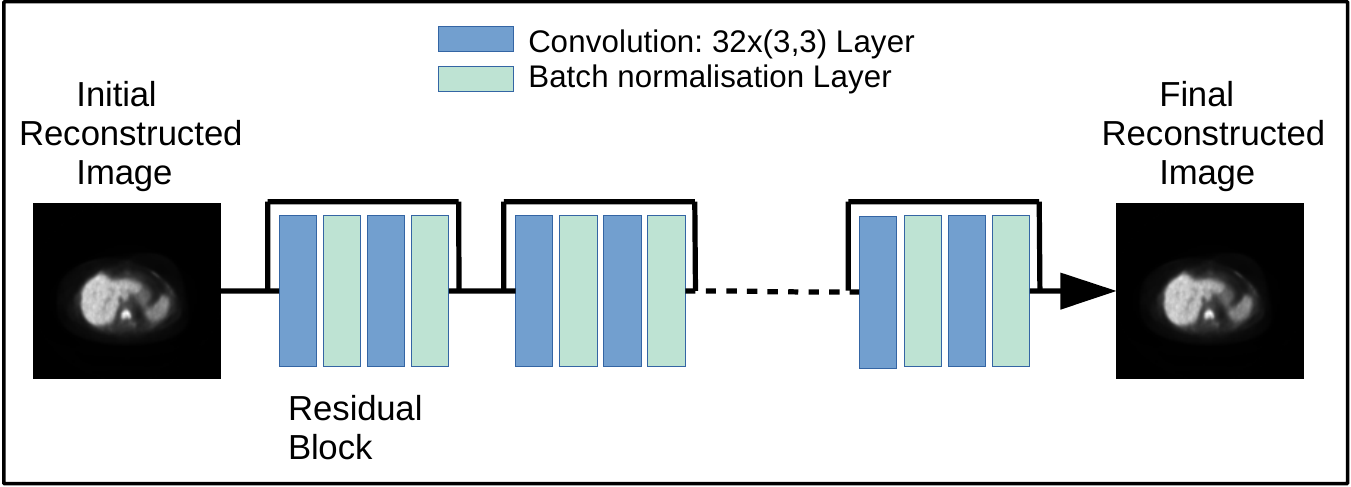}
	\caption{Representation of the super resolution block. It consists of 8 residual blocks with Convolution, Batch normalization and PReLu.}
	\label{fig:super}
\end{figure}

\subsubsection{Super Resolution}
The function of the \ac{SR} is to improve the estimate produced by the image reconstruction network. Several works already exist concerning single image super resolution \cite{ledig2017photo,lim2017enhanced}. In this work we employed a basic super residual network architecture to improve the reconstruction. It consists of convolutional blocks followed by batch normalization with \ac{PReLU} activation. There were a total of 8 residual blocks in the network as represented in Figure~\ref{fig:super}. The loss function used in this architecture was perceptual loss:

\begin{equation}
      \mathrm{Perceptual Loss} =  |\mathrm{VGG}_{16}(\boldx^\star) -\mathrm{VGG}_{16}( \boldxhat)| 
\end{equation}
 $\mathrm{VGG}_{16}(\boldx^\star)$ and $\mathrm{VGG}_{16}( \boldxhat)$ are the extracted features with $\mathrm{VGG}_{16}$ convolutional neural network \cite{simonyan2014very} for the true and predicted image.

The features are extracted from the 10th layer of the VGG architecture i.e., only the first three convolutional blocks are considered. We observed that extracting deeper features led to the network hallucinating features in the reconstructed images.

\subsection{Dataset Description}
\begin{figure}[!htbp]
	\centering
	\includegraphics[width=0.99\linewidth]{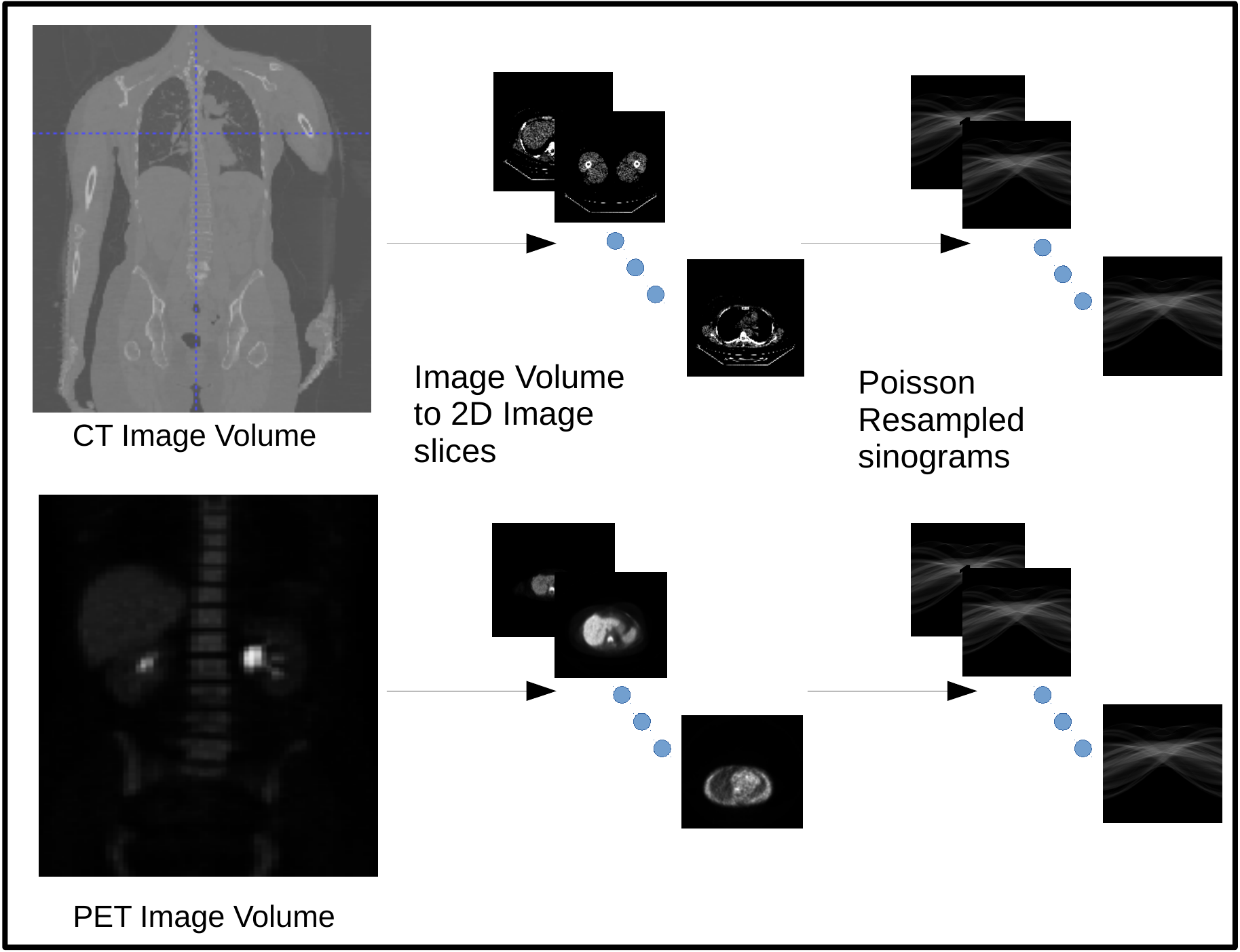}
	\caption{Data preparation}
	\label{fig:data_prep}
\end{figure}

\begin{figure}[!htbp]
	\centering
	\includegraphics[width=0.7\linewidth]{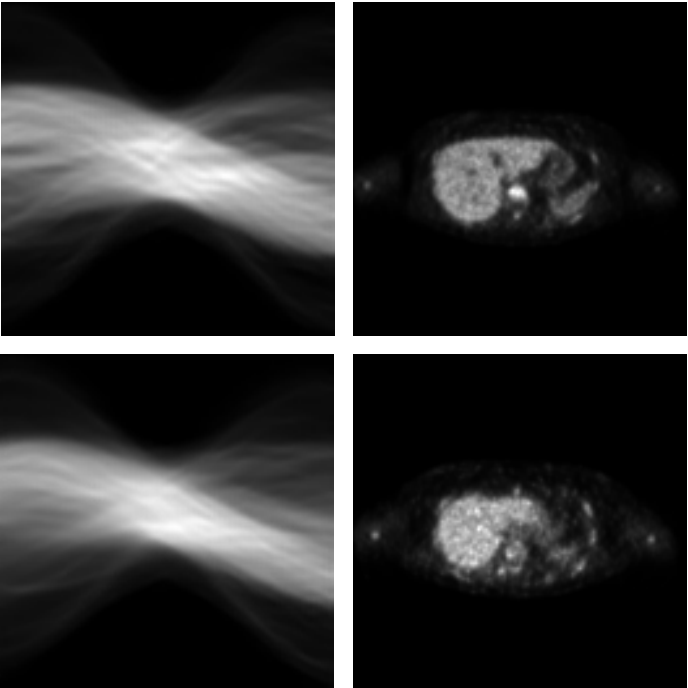}
	\caption{Example PET sinogram-image pairs from the dataset}
	\label{fig:realdata}
\end{figure}

\begin{figure}[!htbp]
	\centering
	\includegraphics[width=0.7\linewidth]{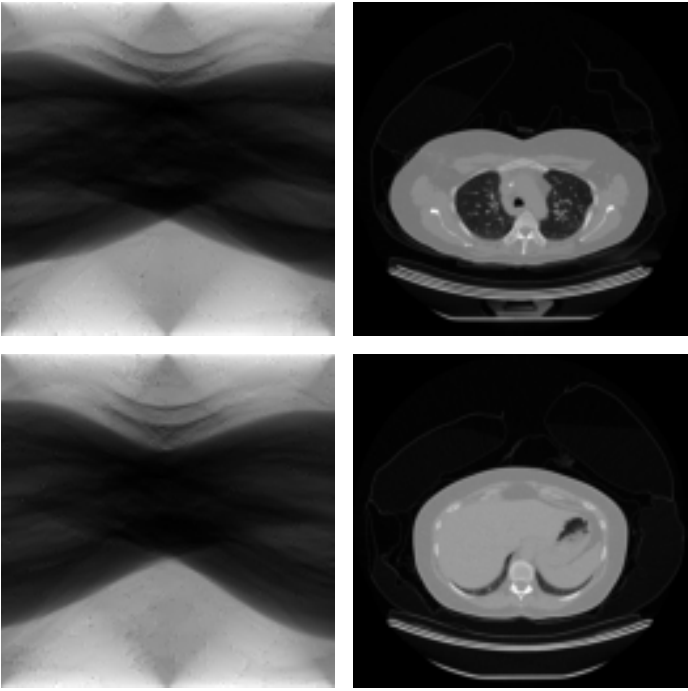}
	\caption{Example CT sinogram-image pairs from the dataset }
	\label{fig:phantom}
\end{figure}

We applied our methodology on fluorothymidine (FLT) PET/CT images from the American College of Radiology Imaging Network (ACRIN) FLT Breast \ac{PET}/\ac{CT} database \cite{kostakoglu2015phase}. The details of the dataset are given in Table~\ref{table:2}. The sinograms were initially generated by projecting 2-D \ac{PET} and \ac{CT} images slices with the Python SKLEARN Radon transform, following the models \eqref{eq:CT} and \eqref{eq:PET} for \ac{CT} and \ac{PET} respectively, with Poisson noise added. 
The methodology represented in Figure~\ref{fig:data_prep} was used for data preparation for the \ac{PET} and \ac{CT} modalities respectively. Sample pairs from the \ac{PET} and \ac{CT} datasets are shown in Figures~\ref{fig:realdata} and \ref{fig:phantom}. The \ac{CT} images were downsized from $512\times{}512$, and the reconstruction was implemented for 2-D $128\times{}128$ images.

\begin{table}[h!]
\caption{Dataset Description}
\label{table:2}
\centering
 \begin{tabular}{||c|c||} 
 \hline
 Dataset Statistics &  \\ [0.5ex] 
 \hline
 Modalities & CT, PET   \\ 
 \hline
 Number of Patients  & 83  \\
 \hline
 Number of \ac{PET} \ac{2D} Image slices & 76,000 \\ 
 \hline
 Number of CT \ac{2D} Image slices & 21,104 \\ 
 \hline
 PET Matrix size & 128 \\ 
 \hline
 CT Matrix size & 512 \\ 
 \hline
 Scanner & GE Discovery ST \\
 \hline
  \end{tabular}

\end{table}


\subsection{Training}
TensorFlow \cite{abadi2016tensorflow} and Keras \cite{chollet2015keras} were used for the realization of the architectures described in the section above. These architectures were implemented on a single Nvidia GeForce GTX 2080Ti GPU. 
A collection of images $\{\boldx^\star_k\}_{k=1}^N$ was used to generate a corresponding collection of noiseless sinograms  $\{\boldy^\star_k\}_{k=1}^N$ following models \eqref{eq:CT} and \eqref{eq:PET}, low-counts and high-counts sinograms, $\{\boldy^\mathrm{LC}_k\}_{k=1}^N$ and $\{\boldy^\mathrm{HC}_k\}_{k=1}^N$ sinograms were generated by adding Poisson noise, where the expected number of counts was adjusted by rescaling the intensity. The denoising segment was trained with the collection $\{(\boldy^\mathrm{LC}_k,(\boldy^\mathrm{HC}_k)\}_{k=1}^N$, using the high-count sinograms as ground truth, with total number of training samples $N$ = 120000 and number of epochs $M$ = 100.  
For the training of the second segment, we used a collection of denoised sinograms, namely  $\{\boldyhat_k\}_{k=1}^N$, and their corresponding ground true images $\{\boldx^\star_k\}_{k=1}^N$. 
The image reconstruction segment was trained according to the Algorithm~\ref{alg:1}, that is to say by alternating between training $G_1$ with $\{(\boldyhat_k, \boldx^\star_k)\}_{k=1}^N$ and training  $G_2$ with $\{(\boldxhat_k,\boldy^\star_k)\}_{k=1}^N$, where $\boldxhat_k$ is a prediction from $G_1$ with $\boldxhat_k$ as an input. This segment was trained with $N = 40000$ and $M = 50$.
The \ac{CT} data were augmented by rotating the data by 90 degrees to generate the required training data. Owing to the larger \ac{PET} dataset it was not necessary to perform data augmentation. Finally the \ac{SR} segment was trained on the images predicted by the \ac{DUGAN} and the \ac{GT} images $\{(\boldxhat_k, \boldx^\star_k)\}_{k=1}^N $ . The \ac{SR} block was trained with $N = 20000$ for 100 epochs.
For the testing, a set of 2000 sinogram-image pairs were used.

\subsection{Quantitative Analysis}
Testing for the aforementioned architectures was done on samples that were not a part of the training data. The metrics used for this analysis are \ac{RMSE} and \ac{SSIM} Index. They are defined below:

\begin{equation}
  \mathrm{RMSE}(\boldx^\star,\boldxhat) = \sqrt{\frac{1}{n}   \sum_{j=1}^{m} (x^\star_j-\xhat_j )^2} 
\end{equation}

where $n$ is the number of pixels. $\xhat$ is the \ac{GT} $x^\star$ the predicted output. \\
\begin{equation}
      \mathrm{SSIM}(\boldx^\star,\boldx) = \frac{(2\mu_{\boldx^\star}\mu_{\boldx}+c_{1})(2\sigma_{\boldx^\star\boldx}+c_{2})}{(\mu_{\boldx^\star}^2+\mu_{\boldx}^2+c_{1})(\sigma_{\boldx^\star}^2+\sigma_{\boldx}^2+c_{2})}   
\end{equation}
where $\mu_{\boldx^\star}$ $\mu_{\boldx}$ are the averages of $\boldx^\star$ and $\boldx$ respectively, $\sigma_{\boldx^\star}^2$ and $\sigma_{\boldx}^2$ are the variances of $\boldx^\star$ and $\boldx$, $\sigma_{\boldx^\star\boldx}$ is the covariance between $\boldx^\star$ and $\boldx$ , $c_{1}=(k_{1}L)^2$ and $c_{2}=(k_{2}L)^2$ where $k_{1}=0.01$ and $k_{2}=0.03$ by default. 
\subsection{Region of Interest Analysis}

The \ac{SNR} and \ac{CNR} were studied for four regions of interest identified within the patient body. The \ac{SNR} and \ac{CNR} were evaluated by treating a region as foreground and the other three regions as background. 
\begin{equation}
\mathrm{SNR} = \frac{\mu_\mathrm{r}-\mu_\mathrm{b}}{\sigma_\mathrm{b}}
\label{SNR}
\end{equation}

\begin{equation}
\mathrm{CNR} = \frac{|{\mu_\mathrm{r}-\mu_\mathrm{b}}|}{\sqrt{\sigma_\mathrm{r}^2+\sigma_\mathrm{b}^2}}.
\label{CNR}
\end{equation}

where $\mu_\mathrm{r}$ and $\mu_\mathrm{b}$, $\sigma_\mathrm{r}$ and $\sigma_\mathrm{b}$ correspond to the mean and standard deviation in the \ac{ROI} and the background respectively. In this study we compared the initial reconstructed output of the \ac{DUGAN}, the final reconstruction along with \ac{SR} and the original \ac{GT} which was reconstructed with GE discovery ST using an \ac{OSEM} algorithm..     

\subsection{Comparison with DeepPET}

We implemented the architecture DeepPET \cite{haeggstroem2018deeprec} and compared the predictions with our proposed approach for the reconstruction of PET images. DeepPET was trained on $\{(\boldyhat_k, \boldx^\star_k)\}_{k=1}^N$, notation similar to the training section from above, with $N$ = 120000, exclusively on \ac{PET} data. It is worth noting that the input and output dimensions in our study are identical while it was not originally for DeepPET. The architecture of DeepPET is summarized in Figure~\ref{fig:super}. This architecture was trained for 100 epochs with an Adam optimiser. 

\begin{figure}[!htbp]
	\centering
	\includegraphics[width=1.0\linewidth]{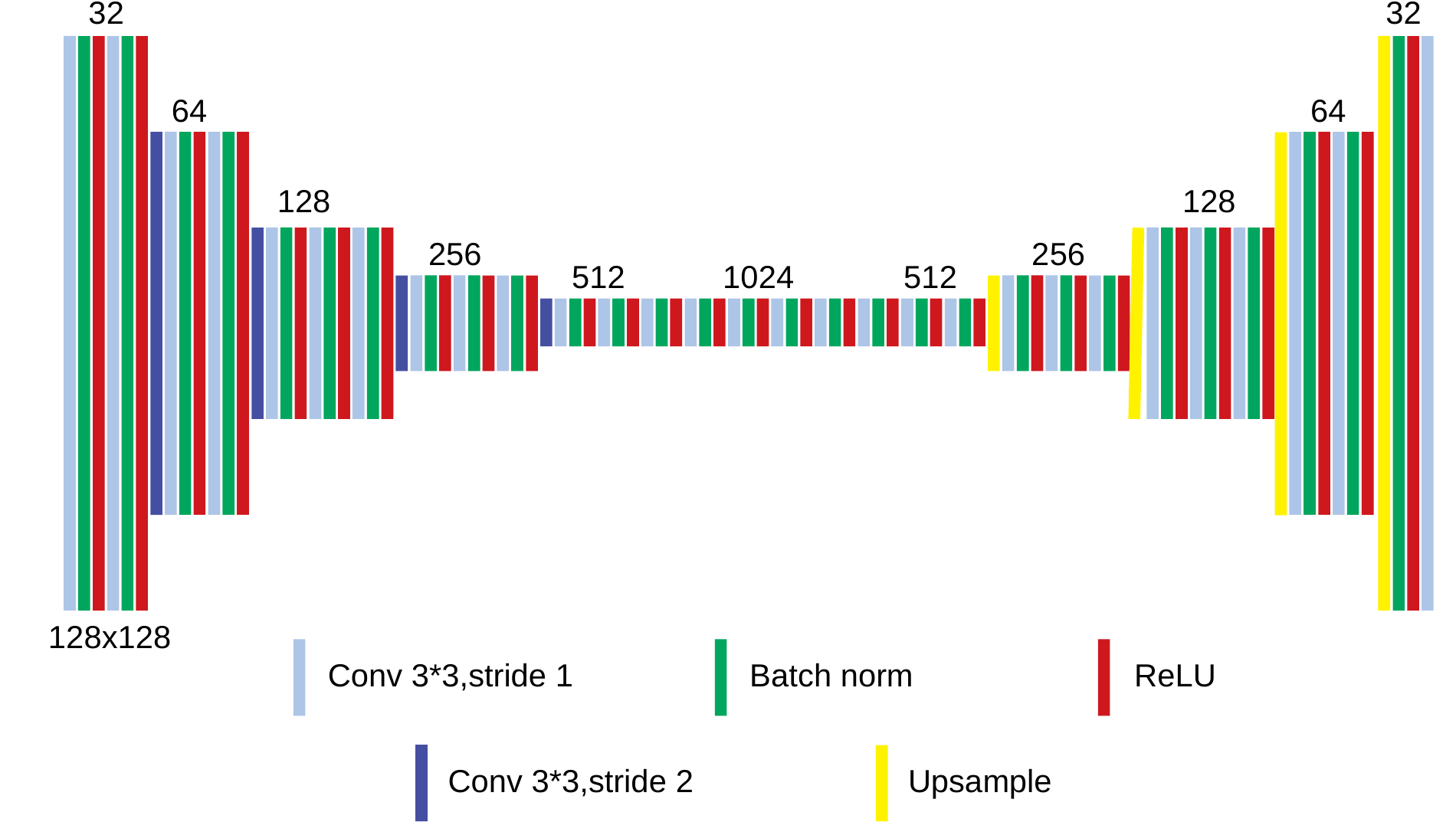}
	\caption{Representation of DeepPET. The number of filters in each convolutional layer is labeled on top of each block.}
	\label{fig:deepPET}
\end{figure}

\section{Results}

\begin{figure*}[!htbp]
	\centering
	\includegraphics[width=0.7\linewidth]{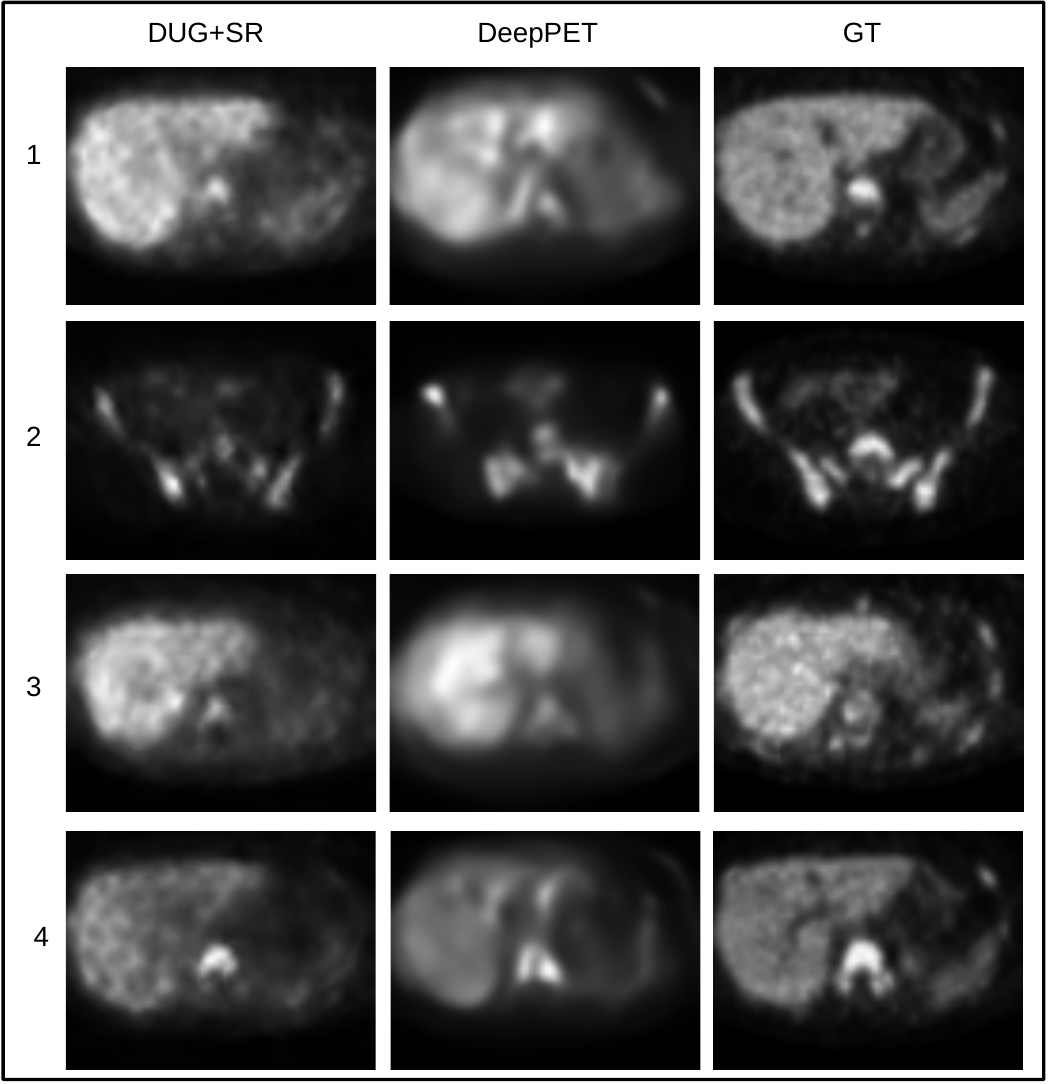}
	\caption{Image predictions by \ac{DUGAN}+SR, DeepPET and \ac{GT} for four \ac{PET} Images from different parts of the patient volume}
	\label{fig:PET_results}
\end{figure*}

\begin{figure}[!htbp]
    \centering
	\includegraphics[width=1.0\linewidth]{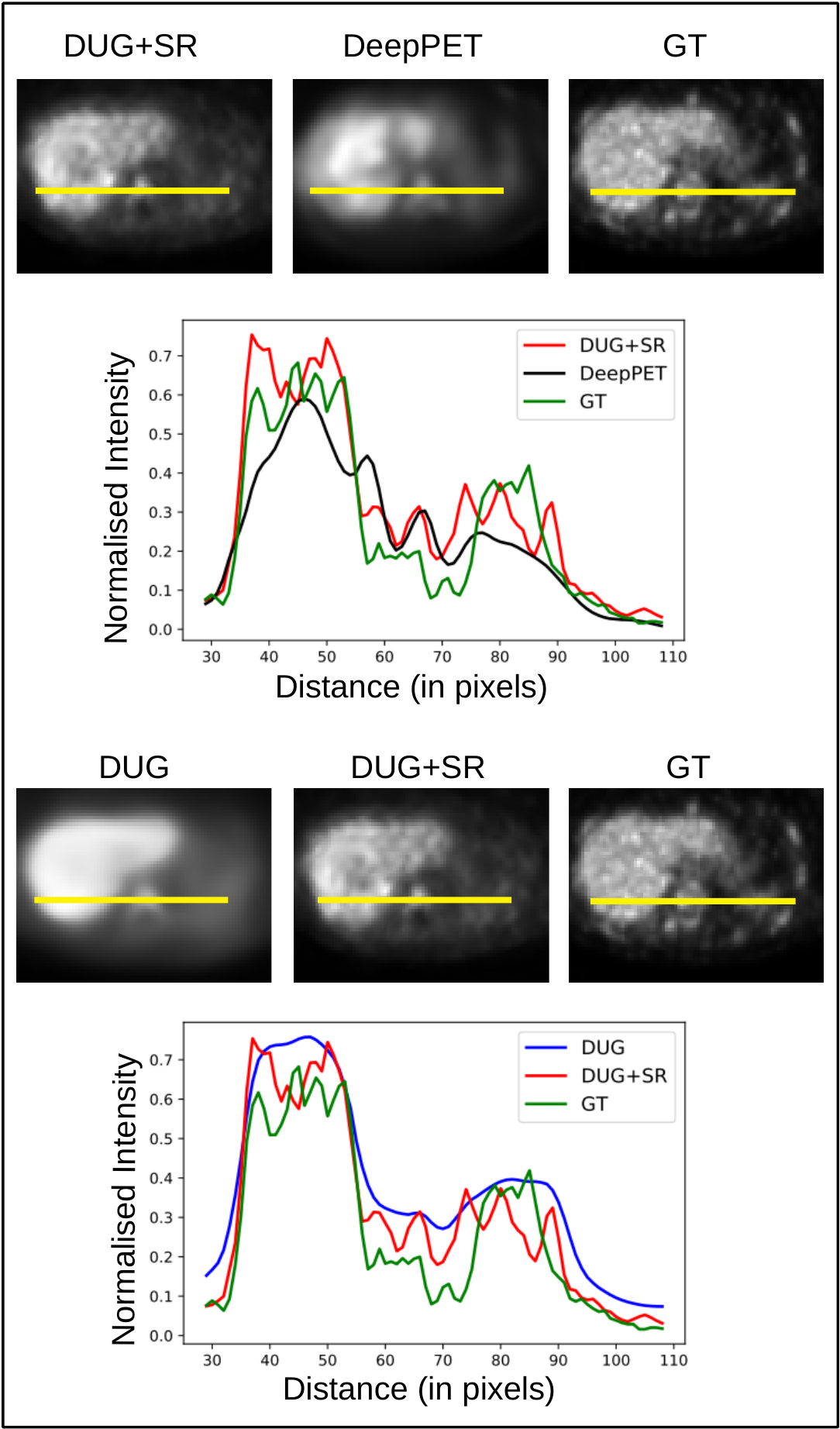}
	\caption{Intensity Profile across the image (highlighted by a yellow line) for a \ac{PET} image prediction by \ac{DUGAN}, \ac{DUGAN}+SR and DeepPET compared with the \ac{GT}}
	\label{fig:PET_ip}
\end{figure}

\begin{figure}[!htbp]
	\centering
	\includegraphics[width=1.0\linewidth]{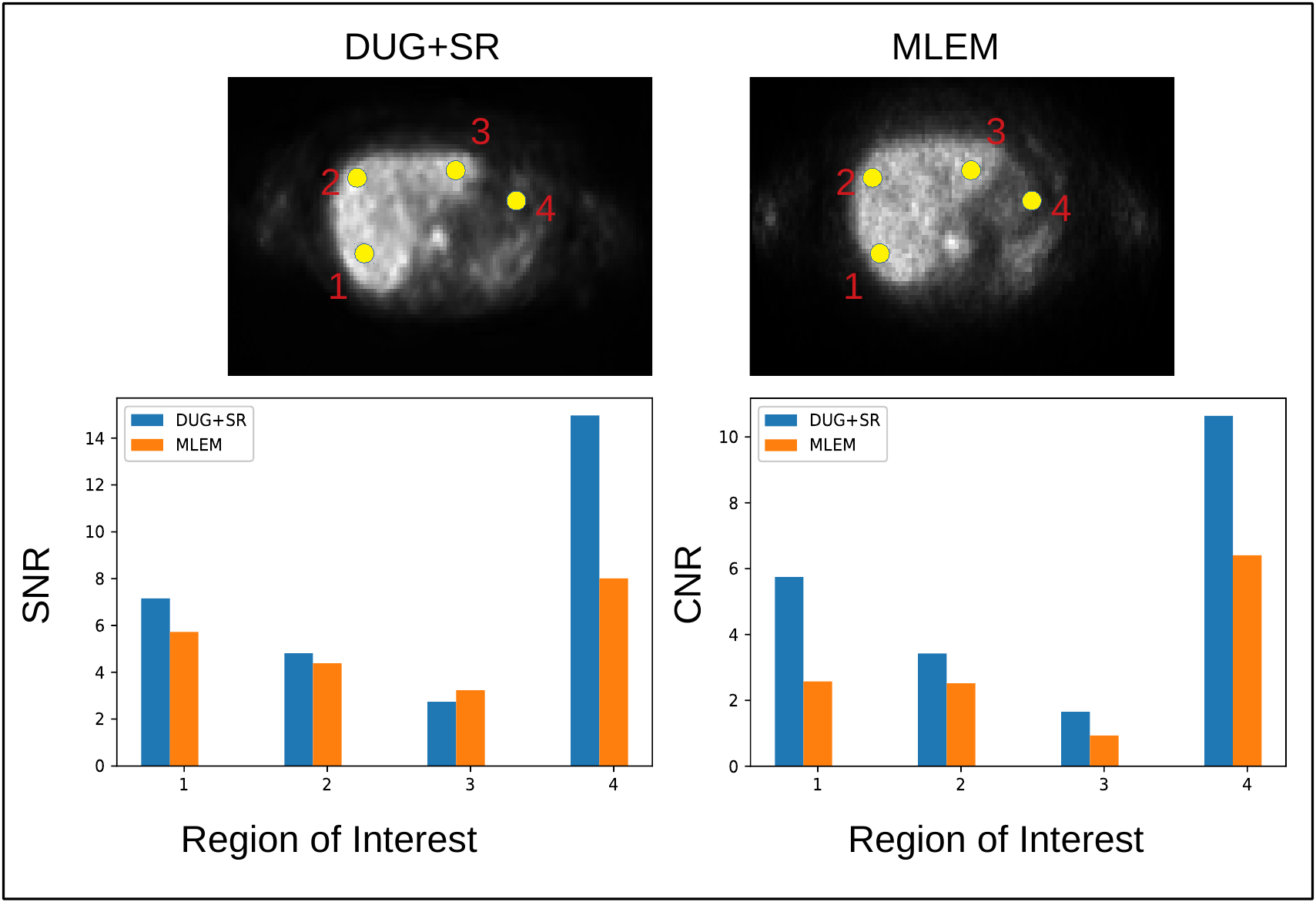}
	\caption{\ac{SNR} and \ac{CNR} comparison amongst DUG+SR and OSEM for \ac{PET} image along 4 regions of interest}
	\label{fig:PET_roi}
\end{figure}

\begin{figure*}[!htbp]
	\centering
	\includegraphics[width=0.8\linewidth]{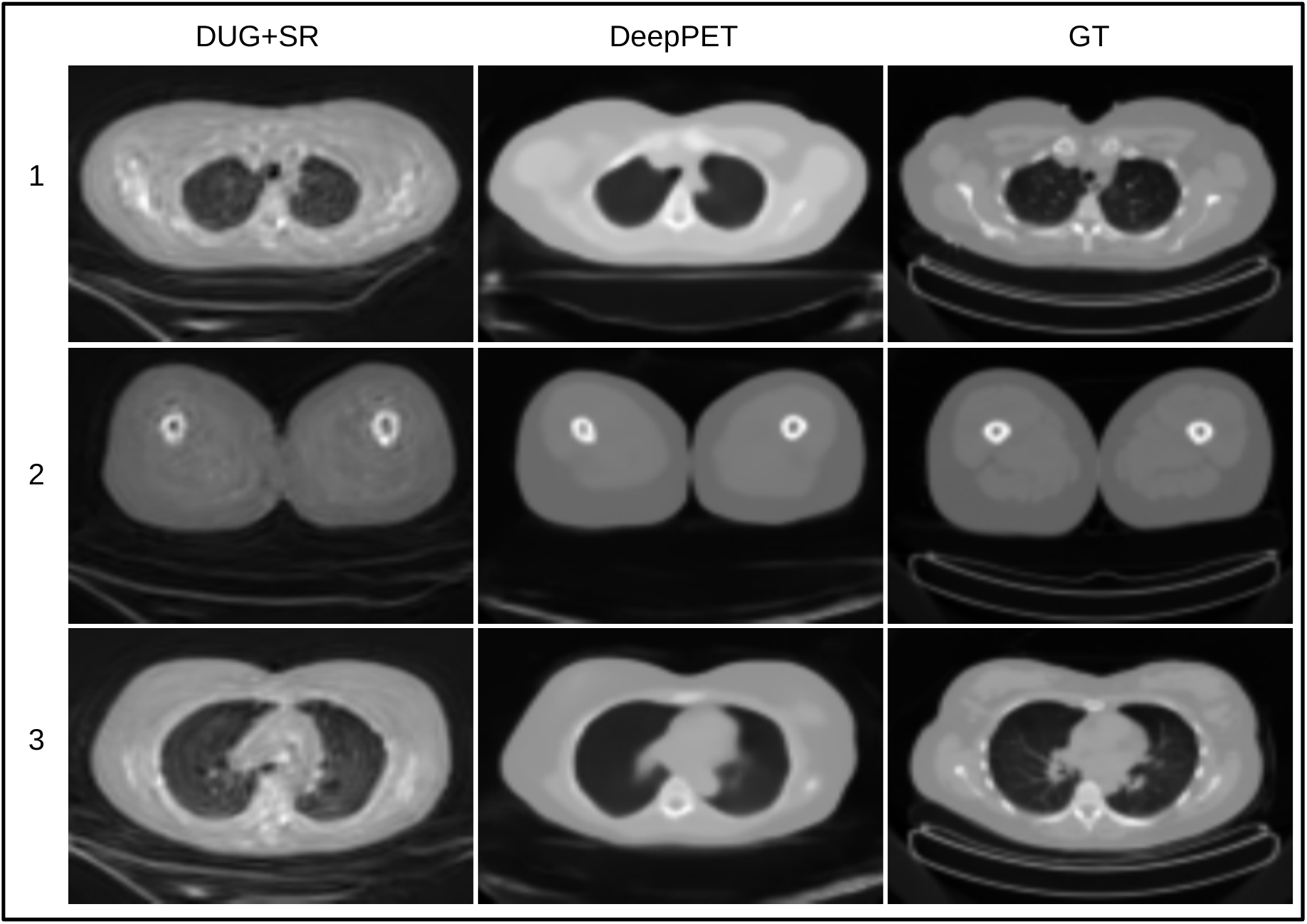}
	\caption{Image predictions by \ac{DUGAN}+ \ac{SR}, DeepPET and \ac{GT} are displayed for 3 \ac{CT} Images along different parts of the patient volume.}
	\label{fig:CT_results}
\end{figure*}

\begin{figure}[!htbp]
    \centering
	\includegraphics[width=1.0\linewidth]{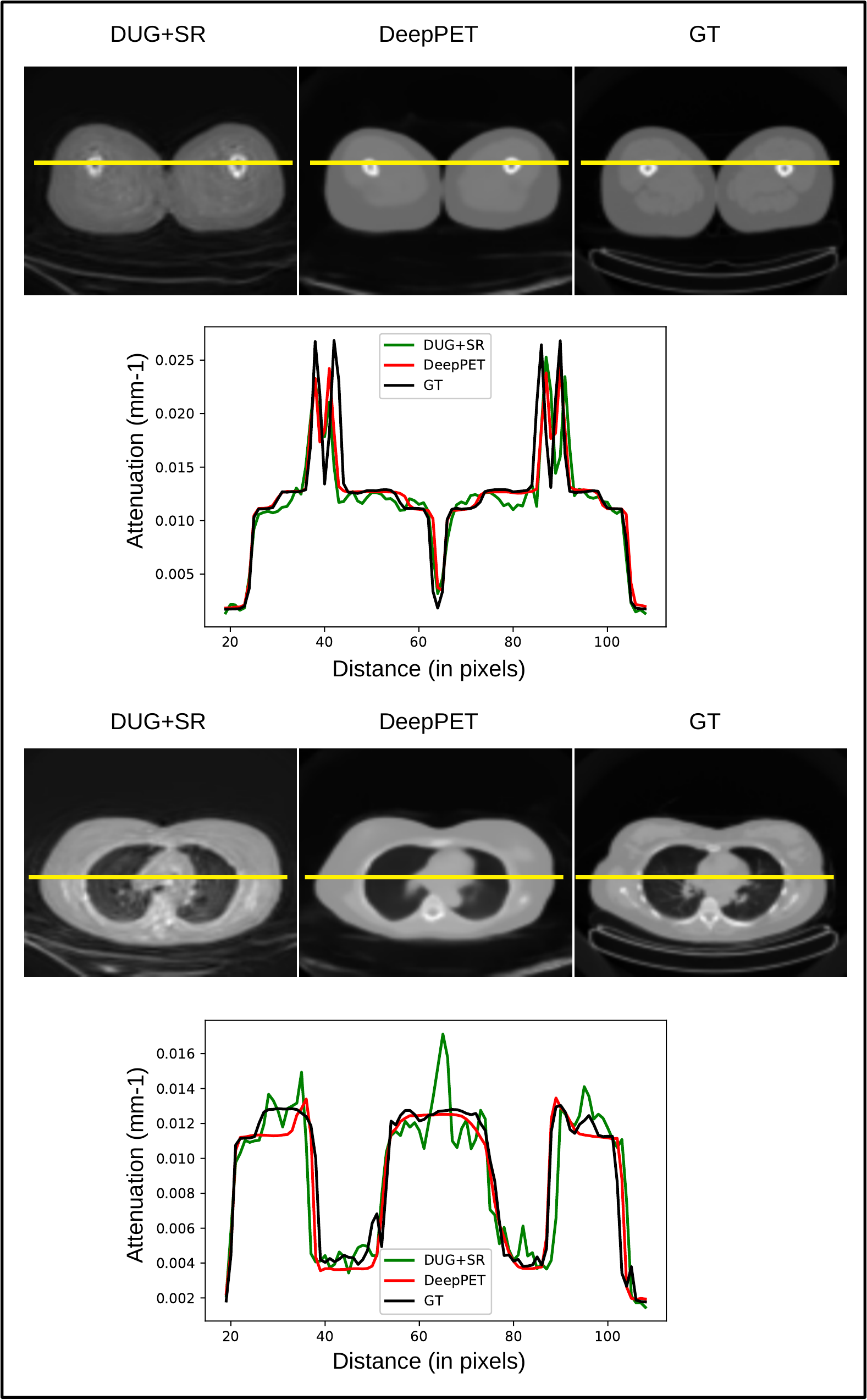}
	\caption{Intensity Profile for two \ac{CT} images (highlighted by a yellow line) predicted by \ac{DUGAN} and \ac{SR} compared with the \ac{GT}}
	\label{fig:CT_ip}
\end{figure}

The predictions from the architectures along with the \ac{GT} and the sinogram are shown in Figure~\ref{fig:PET_results} for \ac{PET} images. The results are displayed for four test image slices across the columns. Each column shows the predicted output by the proposed DUG-RECON architecture and the DeepPET architecture, as well as the \ac{GT}. With regards to the proposed architecture it is observed that the initial reconstructed image i.e., the output of the \ac{DUGAN} looks blurred while final reconstructed output from the super resolution block has noticeably improved details. The predictions by DeepPET are also visibly blurred compared to the final reconstructed output of the proposed architecture and the \ac{GT}. These observations are further ascertained in Table~\ref{table:3} where the quantitative metrics are tabulated. Figure~\ref{fig:PET_ip} provides a comparison of the intensity profiles for the predictions by \ac{DUGAN}, DUG+SR and DeepPET w.r.t. to the \ac{GT} for \ac{PET} images. These intensity values are observed along the line marked in yellow in these figures. As this figure shows, the intensity profile of the final reconstructed image of the proposed architecture is closest to the \ac{GT}. The predictions by \ac{DUGAN} and DeepPET are smoother compared to the predictions by \ac{DUGAN}+SR and the \ac{GT}. The \ac{ROI} analysis is tabulated in Table~\ref{table:5} for the four regions marked in Figure~\ref{fig:PET_roi}. This analysis was carried out for final predictions by the proposed architecture and \ac{MLEM}. Looking closely at Table~\ref{table:5} we notice that the mean values of the deep learning predicted image and the \ac{MLEM} reconstructed image are comparable. The results for \ac{CT} images are displayed in Figure \ref{fig:CT_results}. This Figure provides a comparison between reconstructed image predictions with the proposed architecture, DeepPET with respect to the \ac{GT}. The high-resolution nature of the \ac{CT} images and a smaller dataset presented challenges during the training of the proposed architecture. The predictions by DeepPET appear to be better resolved than those by the proposed architecture. However, the tissue and the bone structures are not clearly seen in the predictions by the deep learning architectures, thereby requiring further work to improve the reconstruction. The intensity plots are compared for two different images in the Figure~\ref{fig:CT_ip}. The \ac{ROI} analysis was carried out for 4 regions marked in the images reconstructed with deep learning and \ac{FBP}. The image reconstructed with \ac{FBP} has better \ac{SNR} and \ac{CNR} compared to the image reconstructed with the proposed architecture.
\begin{table}[h!]
\centering
\caption{The \ac{SSIM} and \ac{RMSE} for the \ac{PET} images are evaluated for 4 different \ac{2D} slices. These metrics are compared amongst predictions by \ac{DUGAN}, DUG+SR and DeepPET}
\label{table:3}
 \begin{tabular}{||c|c|c|c||} 
 \hline
 Image & Architecture & \ac{RMSE} & \ac{SSIM} \\ [0.5ex] 
 \hline\hline
 1 & DUG & 0.059 & 0.74 \\ 
   & DUG+SR & 0.038 & 0.84 \\
   & DeepPET & 0.047 & 0.80 \\
 \hline
 2 & DUG & 0.043 & 0.76 \\ 
   & DUG+SR & 0.046 & 0.86 \\
   & DeepPET & 0.054 & 0.85 \\
 \hline
 3 & DUG & 0.050 & 0.76 \\ 
 
   & DUG+SR & 0.038 & 0.85 \\  
   & DeepPET & 0.043 & 0.83 \\  
 \hline  
 4 & DUG & 0.061 & 0.70 \\ 
    & DUG+SR & 0.045 & 0.82 \\  
    & DeepPET & 0.048 & 0.79 \\  
 \hline  
\end{tabular}

\end{table}

\begin{table}[h!]
\centering
\caption{The \ac{SSIM} and \ac{RMSE} for the \ac{CT} images are evaluated for 4 different \ac{2D} slices. Here the architecture indicates the prediction by \ac{DUGAN} and that of \ac{DUGAN} along with \ac{SR} segment}
\label{table:4}
 \begin{tabular}{||c|c|c|c||} 
 \hline
 Image & Architecture & \ac{RMSE} & \ac{SSIM} \\ [0.5ex] 
 \hline\hline
 1 & DUG & 0.0083 & 0.90 \\ 
   & DUG+SR & 0.0015 & 0.98 \\
   & DeepPET & 0.0012 & 0.99 \\
 \hline
 2 & DUG &  0.0081 & 0.90 \\ 
   & DUG+SR & 0.0015 & 0.99 \\
   & DeepPET & 0.0014 & 0.99 \\
 \hline
 3 & DUG & 0.0015 & 0.91 \\ 
   & DUG+SR & 0.0018 & 0.98 \\
   & DeepPET & 0.0013 & 0.99 \\  
 \hline  
\end{tabular}

\end{table}



\begin{figure}[!htbp]
	\centering
	\includegraphics[width=0.99\linewidth]{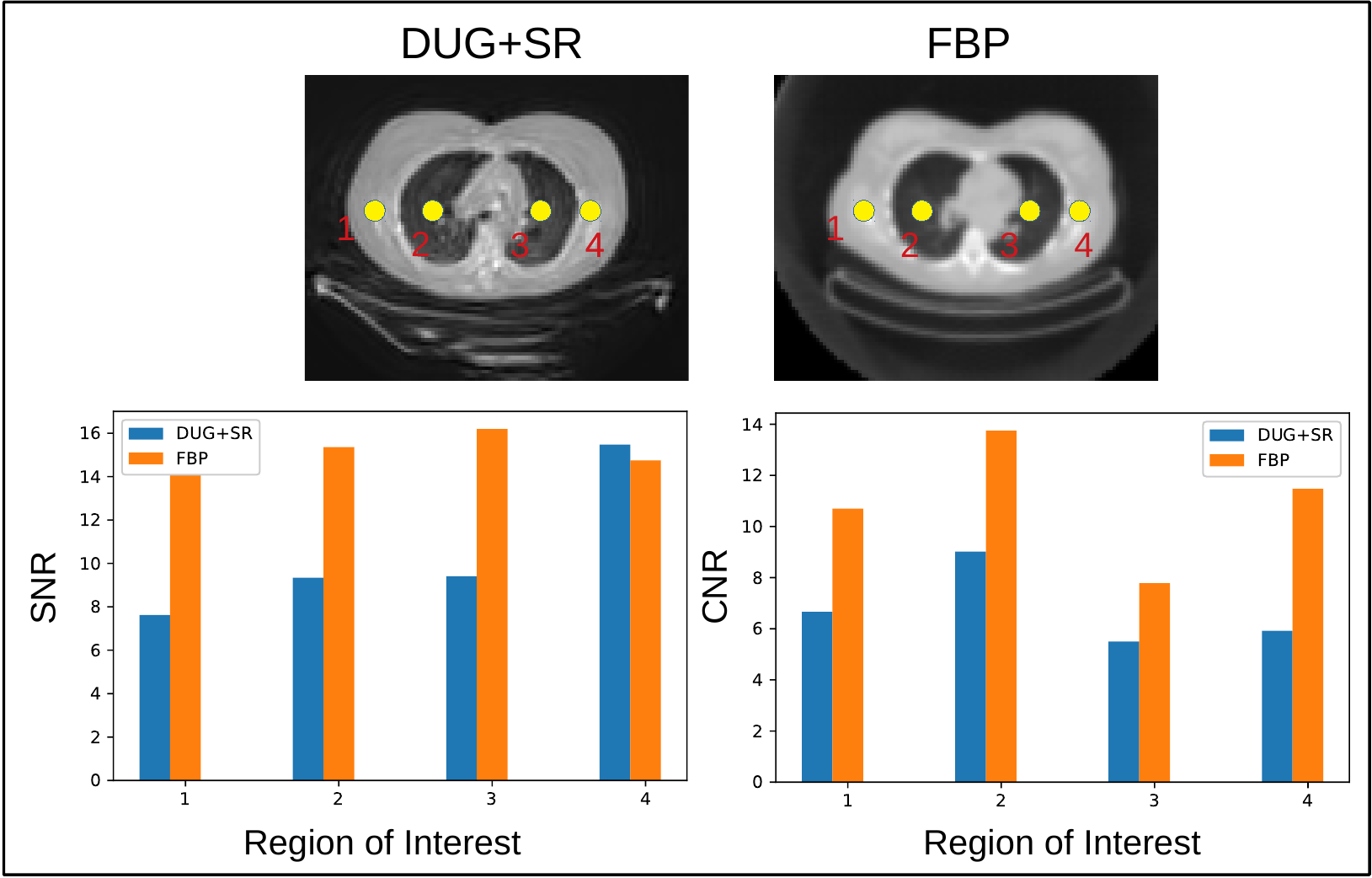}
	\caption{\ac{SNR} and \ac{CNR} comparison amongst DUG+SR and OSEM for \ac{CT} image along 4 regions of interest}
	\label{fig:CT_roi}
\end{figure}
\begin{table}[h!]
\centering
\caption{ROI Analysis: The mean, \ac{SD} and the \ac{SNR} for the 4 regions of interest marked in Figure 12}
\label{table:5}
 \begin{tabular}{||c|c|c|c|c|c||} 
 \hline
 Region & Image & Mean & \ac{SD} & \ac{SNR} & \ac{CNR} \\ [0.5ex] 
 \hline\hline
    1   &    DUG+SR   & 0.706 & 0.024  & 7.15 & 5.71 \\ 
 
       &    MLEM   &    0.676  &  0.035  & 5.72 & 4.55 \\ 
  \hline
    2  &    DUG+SR   & 0.713     &  0.091  & 4.81  & 3.42\\ 
  
       &     MLEM      & 0.648 & 0.11 & 4.38 & 3.26\\ 
  \hline
    3  &    DUG+SR &   0.744  & 0.071 & 2.73 & 1.65\\  
   
       &    MLEM   &    0.547 & 0.154 & 3.23 & 1.22\\ 
 \hline
    4  &    DUG+SR   &    0.117 & 0.008 & 14.96 & 10.64\\  
 
       &    MLEM       &    0.057 & 0.010 & 8.01 & 4.8\\  
 \hline
\end{tabular}

\end{table}

\begin{table}[h!]
\centering
\caption{ROI Analysis: The mean, \ac{SD} and the \ac{SNR} for the 4 regions of interest marked in Figure 15}
\label{table:6}
 \begin{tabular}{||c|c|c|c|c|c||} 
 \hline
 Region & Image & Mean & \ac{SD} & \ac{SNR} & \ac{CNR} \\ [0.5ex] 
 \hline\hline
    1   &    DUG+SR   & 0.011 & 2.91e-4  & 7.61 & 6.66 \\ 
 
       &    FBP   &    0.011  &  3.45e-4  & 15.86 & 10.69 \\ 
  \hline
    2  &    DUG+SR   & 0.004     &  1.53e-4  & 9.34  & 9.02\\ 
  
       &     FBP   & 0.004 &  1.84e-4 & 15.36 & 13.74\\ 
  \hline
    3  &    DUG+SR &   0.005  & 5.90e-4 & 9.40 & 5.49\\  
   
       &    FBP    &    0.005 & 4.88e-4 &  16.19 &  7.78\\ 
 \hline
    4  &    DUG+SR   &    0.012 & 8.319e-4 & 15.47 & 5.92\\  
 
       &    FBP       &    0.012 &  2.736e-4 & 14.74 & 11.47\\  
 \hline
\end{tabular}

\end{table}


\section{Discussion}
 
Deep learning has been applied to different fields of medical imaging. The vast majority of developments concern primarily image processing and analysis/classification tasks. Few works devoted in the field of image reconstruction have been largely concentrated in the use of deep learning within classical tomographic reconstruction algorithms. The main objectives of these works have been an improvement in the speed of convergence and the quality of the successive image estimation within the iterative reconstruction process. The alternative approach involving direct image reconstruction through the use of deep learning approaches to estimate images directly from the use of raw data (sinograms or projections) has been much less explored both for PET and CT. 

Most implementations in direct image reconstruction concern the use of fully connected layers which encode the raw data followed by convolutional layers. In most of the proposed implementations a large number of parameters need to be optimised which reduces the computational burden and overall robustness. In this work we have proposed an original direct image reconstruction deep learning framework based on an architecture inspired by convolutional generative adversarial networks used in image to image translation. The implementation is based on the use of a double U-Net generator (DUG) consisting of two cascaded U-Nets. While the first network transforms the raw data to an image the second one assesses the reconstructed image output of the first network by reiterating the relationship between the reconstructed image and the raw data. Two additional blocks were added; namely a network denoising the raw data prior to their input in the DUG network and a super-resolution block operating on the DUG output image in order to improve it's overall quality.  The proposed network was directly trained on clinical datasets for both PET and CT image reconstruction and its performance was assessed qualitatively and quantitatively. 

Deep neural networks usually result in blurred output.  This fact is clear in the predictions made by the \ac{DUGAN} network. Both the qualitative analysis using the profiles through the reconstructed images and the quantitative metrics  \ac{SSIM} and the \ac{MSE}, demonstrate the improvement of the reconstructed images resulting from the incorporation of the SR block. The qualitative analysis also clearly demonstrates the superiority of the proposed algorithm for direct PET image reconstruction in comparison to alternative approaches such as DeepPET. Finally in the \ac{ROI} analysis we observed that the \ac{SNR} and \ac{CNR} are higher with the deep learning approach for the \ac{PET} images while they are lower than the traditional methods for \ac{CT} images. This is consistent with the observations in the qualitative analysis, where the proposed approach was not able to sufficiently resolve different tissue classes in the resulting reconstructed CT images in comparison with the ground truth. 

One of the potential reasons of the worse performance of DUG-RECON for \ac{CT} reconstruction relative to the superior performance observed for PET image reconstruction may be the lower number of available \ac{CT} images in the training process. This limitation will be addressed as part of future work. Despite the lower performance of the proposed architecture for \ac{CT} images it still presents comparable predictions and opens up avenues for deep learning architectures in tomographic reconstruction. In general, the limitations of a deep learning based reconstruction is the adaptability to new data which is very different from the training sample space. Once a practical methodology is identified, one could have a deep learning pipeline with an ensemble of networks trained on different datasets to perform the reconstruction task.

\section{Conclusion}

We have demonstrated the use of generative convolutional networks for the tomographic image reconstruction task. More specifically we have proposed a new architecture for direct reconstruction that approximates the \ac{2D} reconstruction process. Also we have significantly reduced the parameters required for the domain transform task in image reconstruction. The three-step training pipeline based exclusively on deep learning decentralises the various tasks involved in image reconstruction into denoising, domain transform and super resolution. Various super resolution strategies are currently being explored to improve the reconstructed image. Our proposed strategy for tomographic reconstruction will eventually lead to a network based reconstruction as we continue to improve the framework. Currently it does not perform better than traditional methods in terms of utility metrics but still has the advantage of instantaneous reconstruction and an effective denoising strategy. We plan to extend the work on realistic detector data generated through Monte Carlo simulations in addition to sinograms obtained through Radon transform. We are also working on adapting the architecture to raw detector data. Another important aspect of the data based deep learning approach is that the predictions are limited by the quality of the dataset. It becomes essential to have realistic datasets without compromising on the image quality to improve the training of the neural networks.

\bibliographystyle{IEEEtran}
\bibliography{main}

\end{document}